\documentclass[a4paper,11pt]{article}
\usepackage{pos}
\usepackage{comment}
\usepackage[subrefformat=parens]{subcaption}

\title{Development of HPK Capacitive Coupled LGAD (AC-LGAD) detectors}

\author*[a]{Tomoka Imamura}
\author[a]{Sayuka Kita}
\author[b]{Koji Nakamura}
\author[c]{Kazuhiko Hara}

\affiliation[a]{Graduate School of Science and Technology, University of Tsukuba,\\
  1-1-1 Tennodai, Tsukuba, Ibaraki, 305-8571, Japan}

\affiliation[b]{Institute of Particle and Nuclear Studies, High Energy Research Organization,\\
1-1 Oho, Tsukuba, Ibaraki, 305-0801, Japan}

\affiliation[c]{Faculty of Pure and Applied Science and Tomonaga Center for the History of the Universe, University of Tsukuba,\\
1-1-1 Tennodai, Tsukuba, Ibaraki, 305-8571, Japan}

\emailAdd{tomoka.imamura@cern.ch}
\emailAdd{Koji.Nakamura@cern.ch}

\abstract{The detectors with $\mathcal{O}$(10) \textmu m spatial resolution and $\mathcal{O}$(10) ps timing resolution construct powerful particle trackers for future hadron or lepton collider experiments. LGAD: Low-Gain-Avalanche-Diode is a semiconductor detector technology to improve timing resolution.
Capacitive Coupled LGAD (AC-LGAD) detectors have been developed with HPK in order to meet both spatial and timing resolution requirements. Prototype samples with finely segmented electrodes have been produced and tested with various sensor fabrication parameters: doping concentrations, active thickness and electrode coupling capacitance. Timing resolution and signal height were evaluated with beta-ray. As a result, 100 \textmu m pitch pixel detector has been successfully developed achieving a good signal to noise ratio and 30 ps timing resolution for beta-ray. The detectors have to meet radiation hardness requirements as well. Radiation hardness of LGAD detectors has to be improved to use the detectors as inner trackers for hadron colliders. One of the major mechanisms of radiation damage of LGAD detectors is acceptor removal: shallow dopants in the gain layer of LGAD detectors are reduced by radiation damage. Two novel ideas are tested on effectiveness of delaying the acceptor removal.
}

\FullConference{The 32nd International Workshop on Vertex Detectors (VERTEX2023)\\
 16-20 October 2023\\
Sestri Levante, Genova, Italy\\}

\begin{document}
\maketitle

\section{Introduction}
\label{sec:Introduction}
The detectors with $\mathcal{O}$(10) \textmu m spatial resolution and $\mathcal{O}$(10) ps timing resolution are innovative in future hadron or lepton collider experiments in order to deal with the pile-up problem as well as particle identification.
The Low-Gain-Avalanche-Diode (LGAD) is a semiconductor detector technology to improve timing resolution. 
Capacitive Coupled LGAD (AC-LGAD) detectors have been successfully developed~\cite{nim}, achieving a timing resolution of 20 ps for minimum ionizing particles (MIPs). 
As regards spatial resolution, 100 \textmu m pitch pixel detector has been successfully developed with a good signal to noise ratio and small cross talk~\cite{vtxprc22}.\\
\indent The detectors have to meet radiation hardness requirements in order to install it to inner tracking detector for hadron colliders. One of the major problems of radiation hardness of LGAD detector is acceptor removal, where shallow dopants in $p^+$ gain layer is reduced by radiation damage. This requires the detector to be applied higher voltages than before irradiation to compensate the weaker electric field in gain layer, which puts the detector at the risk of single-event-burnout~\cite{SEB}. The prototype detectors adopted with two novel ideas to reduce acceptor removal effect have been produced and tested.\\ 
\indent In this paper, following two things are discussed. First, the timing resolution performance of AC-LGAD detector. Second, the performance of LGAD detectors employing two ideas to mitigate the acceptor removal effects after proton irradiation.
\section{AC-LGAD prototype}
\label{sec:AC-LGAD}
\begin{figure}[b]
    \begin{minipage}[h]{0.5\linewidth}
      \centering
      \includegraphics[keepaspectratio, scale=0.3]{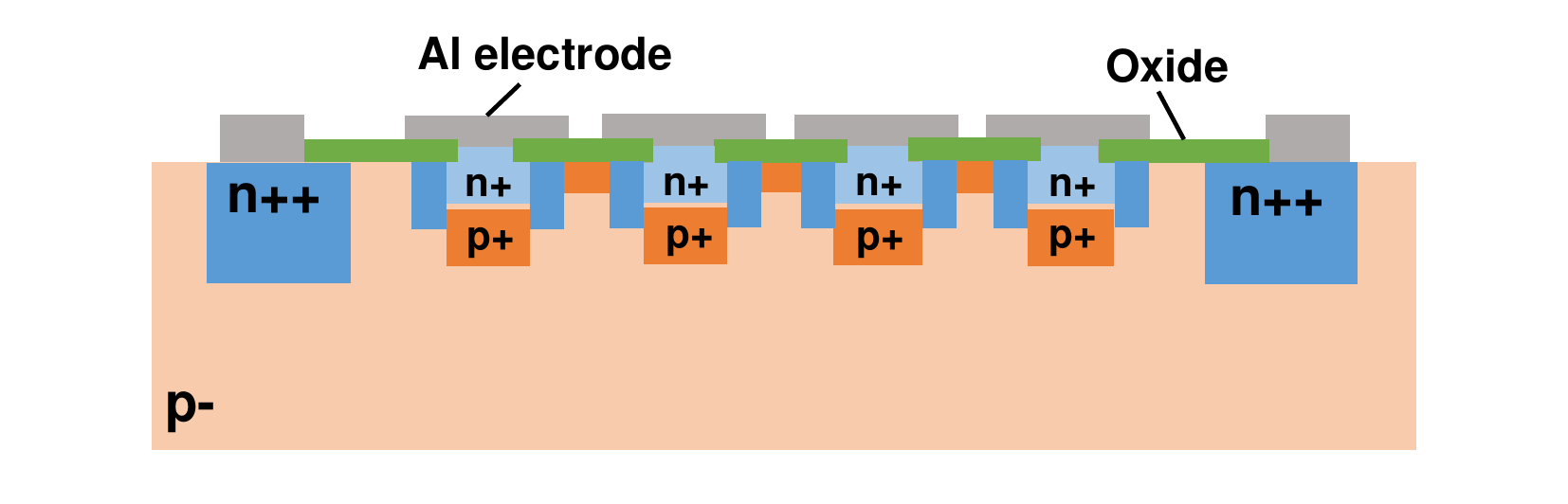}
      \caption{DC-LGAD}
      \label{fig:DC-LGAD}
    \end{minipage}
    \begin{minipage}[h]{0.5\linewidth}
      \centering
      \includegraphics[keepaspectratio, scale=0.3]{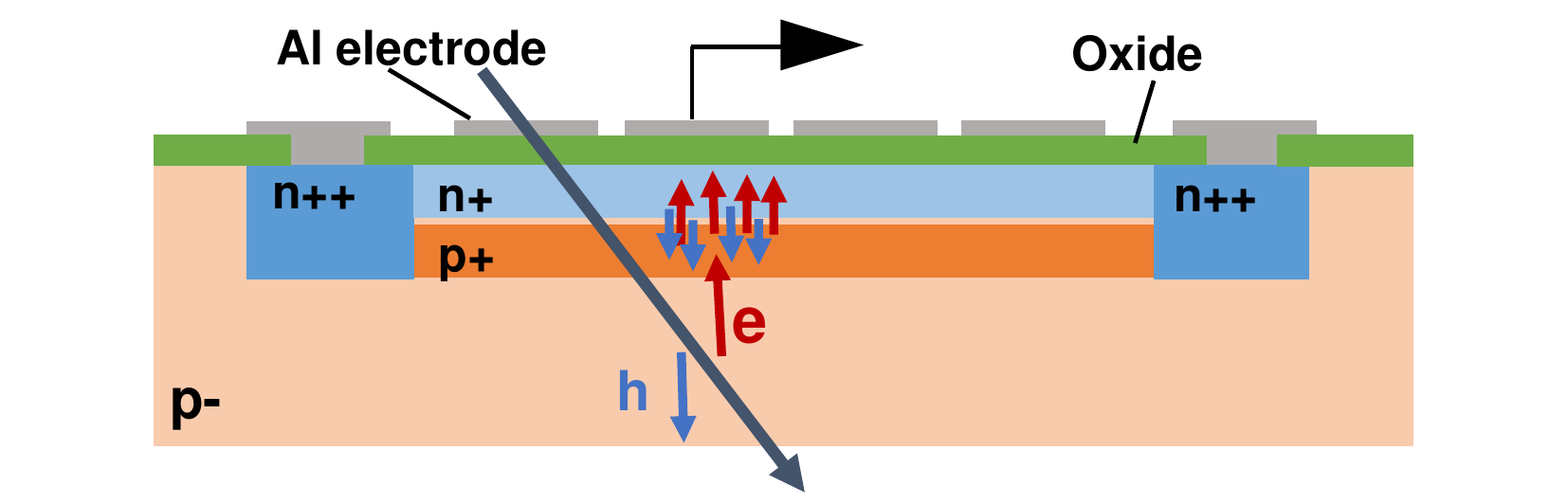}
      \caption{AC-LGAD}
      \label{fig:AC-LGAD}
    \end{minipage}
\end{figure}

The LGAD sensor is an $n^{+}$-in-$p$ semiconductor detector with a gain layer made by implantation of additional $p^{+}$ below the $n^{+}$ implantation. That gain layer makes a local high electric field which develops avalanche multiplication, resulting in generation of large amount of electron and hole pairs. The rapid movement of these electron and hole pairs in the local high electric field creates a large signal instantaneously, leading to a superior timing resolution.\\
\indent In conventional LGAD detector, shown in Fig.~\ref{fig:DC-LGAD}, individual gain layer for each electrode was structured with finely segmented readout. This detector had areas without gain layers between the electrodes, which led to a low fill factor. A capacitively-coupled LGAD (AC-LGAD)  shown in Fig.~\ref{fig:AC-LGAD}, has been developed to solve this low fill factor problem. AC-LGAD has an uniform single gain layer under segmented electrodes which coupled with the gain layer via oxide. AC-LGAD has been successfully developed with pixel pitch 100 \textmu m with 100\% fill factor and small cross talk~\cite{vtxprc22}.\\
\section{Measurement setup}
\label{sec:setup}

The sensor performance is evaluated by using $^{90}$Sr $\beta$ ray. The sensor electrodes are connected to a 16-ch amplifier board~\cite{nim} by wire bonding. The reverse bias voltage is applied to the backside of the sensor via conductive tape. Signal outputs from the amplifier are recorded by LeCroy waverunner 8000HD oscilloscope. The amplifier board with LGAD sensor is set horizontally in a thermostatic chamber and a $^{90}$Sr source placed above the sensor. The Photek MCP-PMT 240 placed under the sensor provides the timing reference and trigger for oscilloscope.

\section{Timing resolution performance}
\label{sec:Timing resolution}
The timing resolution of LGAD sensors consists of three factors, difference of arrival time due to the signal size (time walk), electronics noise (jitter) and the effect of non-uniform charge deposition through the depth by MIP particle (charge collection noise). The jitter is reduced by larger signal and smaller noise. The thinner active thickness sensors performed smaller the charge collection noise by smaller Landau fluctuation of energy deposition. 


\indent Timing resolution is evaluated as the sigma of a gaussian function fitted to the differences between the signal arrival times and timing reference. The charge collection noise effect can be assessed by comparing the timing resolutions obtained 
from $\beta$ ray measurements. Table~\ref{tab:Timing resolution} shows measurement results of timing resolution for each active thickness sample and estimated charge collection noise by calculating jitter from rise time of pulse, signal size and electrical noise. A timing resolution of $\sim$ 30 ps was obtained. The sensor type shown in the table is 2x2 pixelated with pixel size of (500~\textmu m)$^2$. Thinner sensor showed overall better timing resolution due to the suppression of charge collection noise effect while the jitter contribution is the largest due to smaller signal magnitude.

\begin{table}[h]
  \centering
    \begin{tabular}{c|c|c|c}
    \hline
    Sensor active thickness & 50~\textmu m & 30~\textmu m & 20~\textmu m \\
    \hline \hline
    timing resolution [ps] & 38.8 & 31.5 & 31.2\\
    jitter [ps] & 9.8 & 11.8 & 15.9\\
    charge collection noise [ps] & 37.5 & 29.2 & 26.8\\
    \hline
  \end{tabular}
\caption{Timing resolution summary}
\label{tab:Timing resolution}
\end{table}

\section{Radiation tolerance of LGAD}
\label{sec:radiation}
Shallow dopants in gain layer of LGAD are transformed into defect complexities which no longer have characteristics of shallow dopants due to NIEL effect. This transformation of acceptor, acceptor removal, results in increasing of operation voltage~\cite{vtxprc19}. The suppression of the acceptor removal effect is needed to have radiation tolerance of gain layer for the future high energy hadron collider experiments. The prototypes of LGAD samples adopting two novel ideas described in Section~\ref{sec:comp} and Section~\ref{sec:PAB} for suppressing acceptor removal effect were produced, irradiated and evaluated. Irradiation campaign was done at CYRIC at Tohoku University with 70 MeV proton beam under the environment at -15${}^\circ$C. Sensors are irradiated to $ 8\times10^{13}, 6\times10^{14}, 3\times10^{15}\,\rm{n_{eq}/cm^{2}}$ by uniform scanning and tested by $\beta$ ray measurement setup described in Sec \ref{sec:setup}. 

\subsection{Compensation method}
\label{sec:comp}
In the compensation method, both $p^{+}$ and n$^{+}$ are implanted in $p^{+}$ gain layer and structure the $p^+$ layer by the effective $p^+$ which is the difference of $p^+$ and $n^+$. Both $n^{+}$ and $p^{+}$ are reduced by radiation damage due to acceptor and donor removal. If $n^{+}$ removal is faster than $p^+$ removal, effective $p^{+}$ may reduce slower than conventional $p^+$ layer. Prototypes were produced with 5 parameters as shown in Table.~\ref{tab:Compensation}.  In the table, parameter $a$ represents the doping concentration of Reference, for instance, 10 times higher doping concentration of $p^{+}$ in 10B+9.2P than in Reference.
 
\begin{table}[h]
  \centering
  \begin{tabular}{c|c|c|c|c|c}
    \hline
    & 10B+9.2P & 5B+4.05P & 2.5B+1.5P & 1.5P+0.55P & Reference\\
    \hline \hline
    $p^{+}$ Boron & 10$a$ & 5$a$ & 2.5$a$ & 1.5$a$ & $a$\\
    $n^{+}$ Phosphorous & 9.2$a$ & 4.05$a$ & 1.5$a$ & 0.55$a$ & 0 \\
    effective $p^{+}$ & 0.8$a$ & 0.95$a$ & $a$ & 0.95$a$ & $a$\\
    \hline
  \end{tabular}
  \caption{Parameters of compensation prototypes}
  \label{tab:Compensation}
\end{table}

\begin{figure}[b]
     \centering
     \begin{subfigure}[b]{0.3\textwidth}
         \centering
         \includegraphics[width=\textwidth]{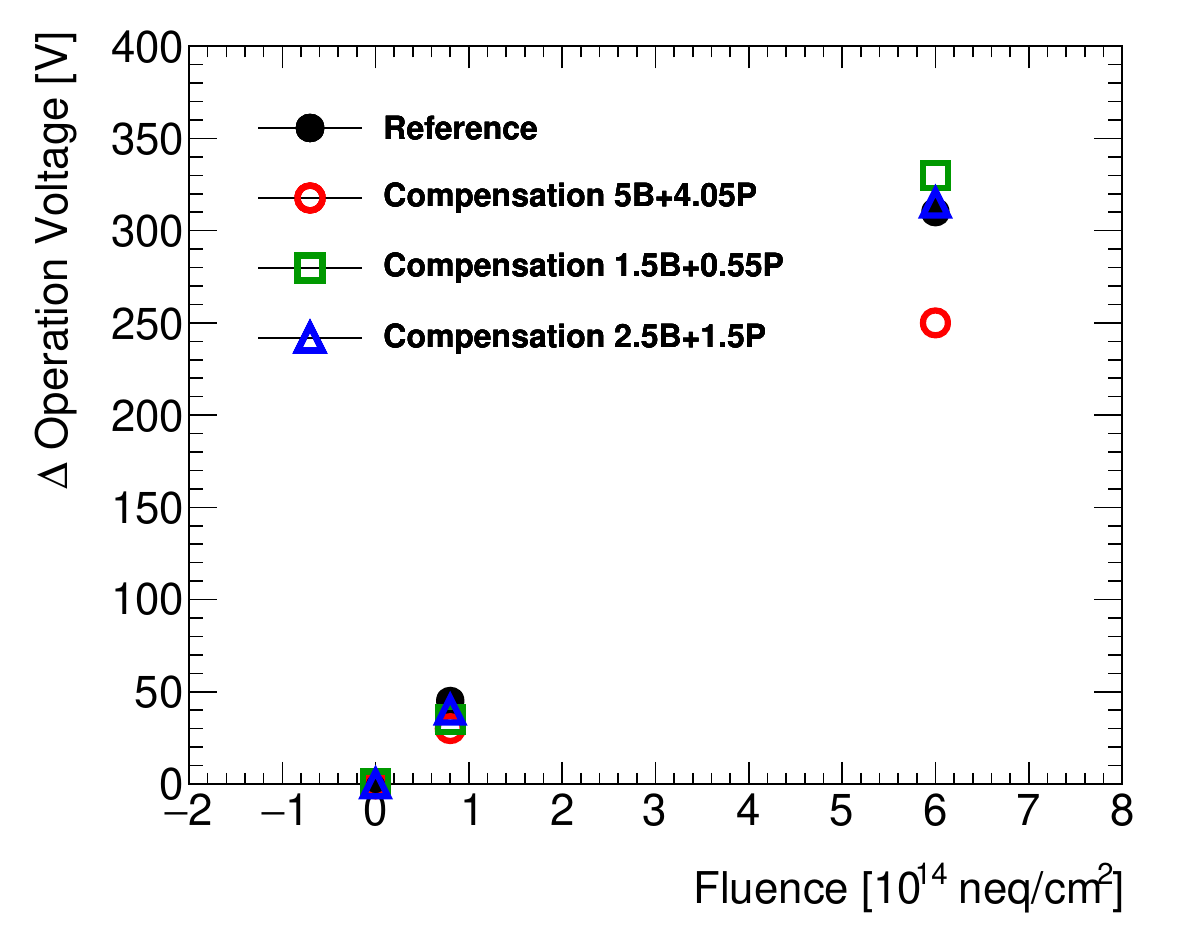}
         \caption{Fluence dependence of increment of operation voltage}
         \label{fig:vopcomp}
     \end{subfigure}
     \hfill
     \begin{subfigure}[b]{0.3\textwidth}
         \centering
         \includegraphics[width=\textwidth]{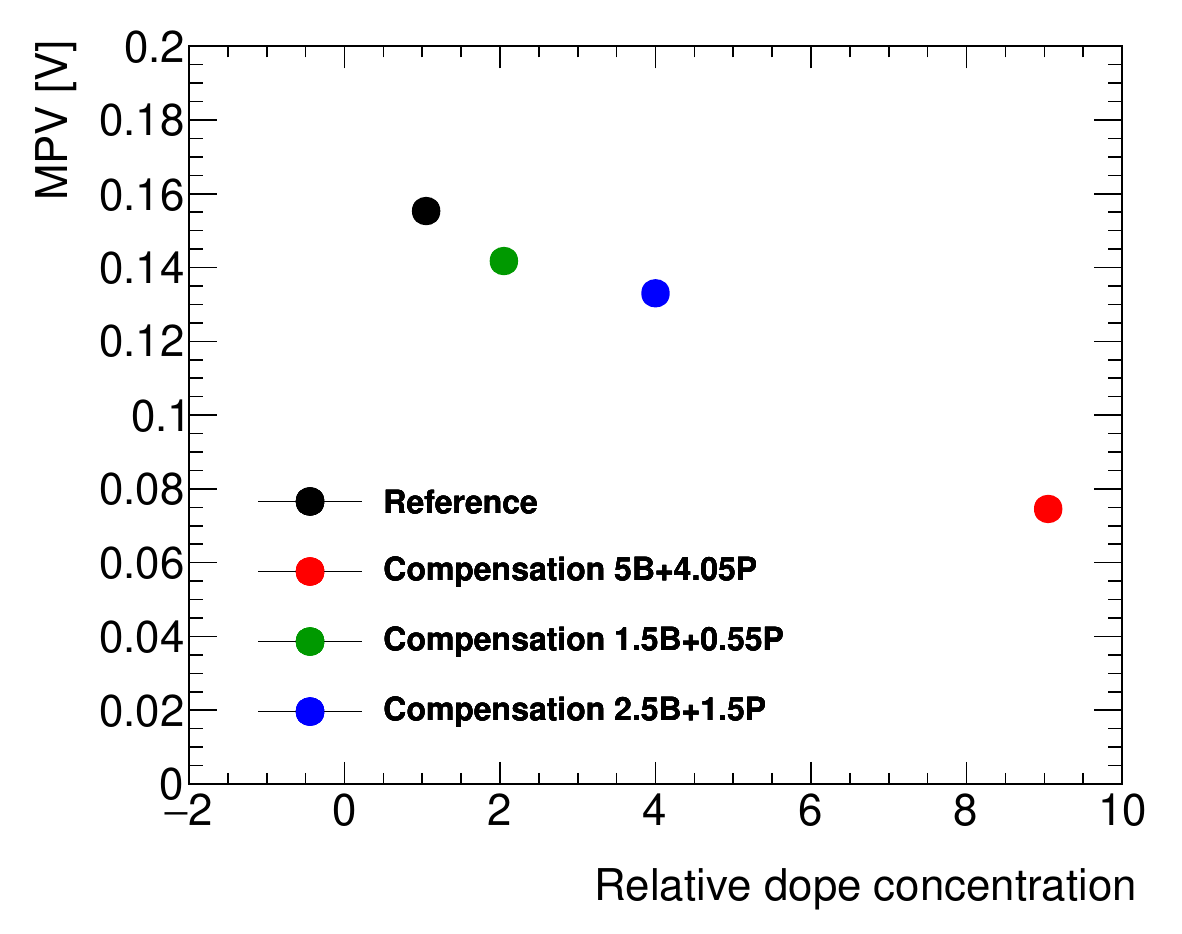}
         \caption{Doping concentration dependence of MPV}
         \label{fig:mpvcomp}
     \end{subfigure}
     \hfill
     \begin{subfigure}[b]{0.3\textwidth}
         \centering
         \includegraphics[width=\textwidth]{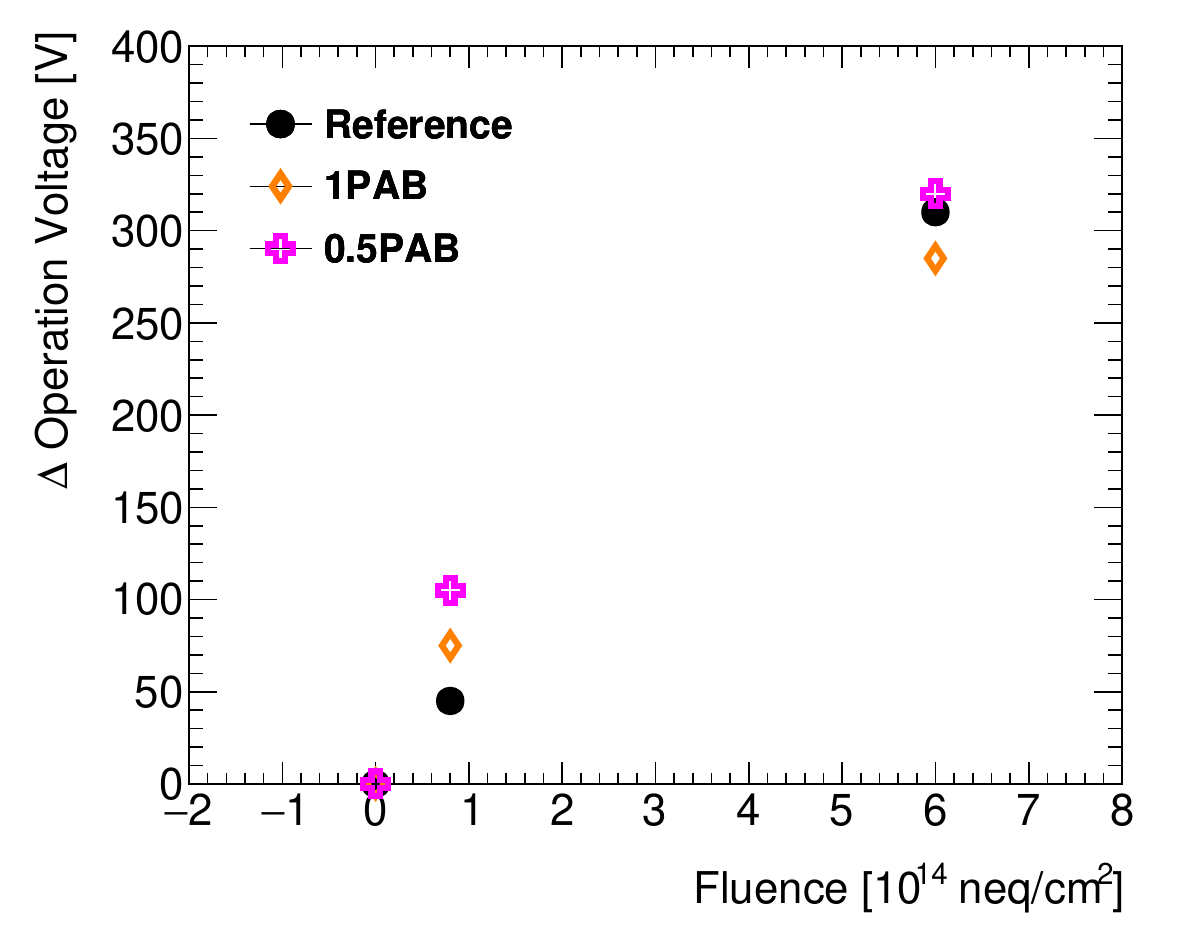}
         \caption{Fluence dependence of increment of operation voltage}
         \label{fig:dvoppab}
     \end{subfigure}
        \caption{Results of compensation prototypes and Partially-Activated-Boron prototypes}
\end{figure}

Sensors were irradiated to the some fluence described in Sec.~\ref{sec:radiation}. Performance of prototypes was evaluated by IV measurement and $\beta$-ray signal measurement before and after irradiation. The fluence dependence of increment of operation voltage is shown in Fig. \ref{fig:vopcomp}. An improvement of suppressing increment has been observed for compensation 5B+4.05P prototype. However, the pulse height (MPV) degradation for higher dope concentration was found as shown in Fig.~\ref{fig:mpvcomp}. It is suspected that avalanche multiplication is obstructed by the dense dopants. Therefore, increase of doping concentration does not simply improve the radiation tolerance.

\subsection{Partially-Activated-Boron}
\label{sec:PAB}
Boron in $p^{+}$ gain layer ais fully activated by a proper annealing process in conventional LGAD sensors, de-activated by NIEL damage and combines with residual oxygen that acts as new donor level~\cite{BiO}. Intrinsic inactive boron before radiation damage cleans up oxygen contamination and this prevents from making additional donors.

The prototypes were produced under two conditions depending on the amount of inactivated boron, 1PAB for adding the same amount of activated boron and 0.5PAB for adding the half amount of activated boron. Sensors were irradiated to the some fluence described in Sec.~\ref{sec:radiation}. Performance of prototypes was evaluated by IV measurement and $\beta$-ray signal measurement with setup described in Sec.~\ref{sec:setup} before and after irradiation. Fluence dependence of increment of operation voltage is shown in Fig.~\ref{fig:dvoppab}. Improvement of suppressing increment of operation voltage was not evident in any of the two conditions.

\section{Conclusion}
AC-LGAD sensors have been successfully developed. Sensors with different thickness down to 20 \textmu m have been tested to reduce charge collection noise. 20 \textmu m prototype achieved 31.2 ps timing resolution for MIPs. To enhance radiation hardness of LGADs, two novel ideas have been tested. Compensation idea is effective with higher doping concentrations but the pulse height before irradiation observed to decrease with the concentration. No significant improvement have been observed for Partially-Activated-Boron idea.

\section*{Acknowledgements}
We would like to acknowledge Hamamatsu Photonics K.K. for the fabrication of various types of AC-LGAD sensors and the discussions with the personnel have been very inspiring and fruitful. This research was partially supported by Grant-in-Aid for scientific research on advanced basic research (Grant No. 19H05193, 19H04393, 21H0073 and 21H01099) from the Ministry of Education, Culture, Sports, Science and Technology, of Japan as well as the Proposals for the U.S.-Japan Science and Technology Cooperation Program in High Energy Physics from JFY2019 to JFY2023 granted by High Energy Accelerator Research Organization (KEK) and Fermi National Accelerator Laboratory (FNAL). In conducting the present research program, the following facilities have been very important:  Cyclotron Radio Isotope Center (CYRIC) at Tohoku University.

\end{document}